\documentclass[10pt]{iopart}
\usepackage{iopams}  
\usepackage{graphicx}
\begin{document}

\title[SPINVERT paramagnetic structure refinement]{SPINVERT: A program for refinement of paramagnetic diffuse scattering
data}

\author{Joseph A. M. Paddison$^{1,2}$, J. Ross Stewart$^2$, and Andrew L. Goodwin$^{1,\ast}$}

\address{$^1$ Department of Chemistry, Inorganic Chemistry Laboratory, University of Oxford,
South Parks Road, Oxford OX1 3QR, U.K.}
\ead{andrew.goodwin@chem.ox.ac.uk}
\address{$^2$ ISIS Facility, Rutherford Appleton Laboratory, Harwell Science and 
Innovation Campus, Didcot, Oxfordshire OX11 0QX, U.K.}

\begin{abstract}
We present a program (SPINVERT; \emph{http://spinvert.chem.ox.ac.uk})
for refinement of magnetic diffuse scattering data for frustrated
magnets, spin liquids, spin glasses, and other magnetically disordered
materials. The approach uses reverse Monte Carlo refinement to fit
a large configuration of spins to experimental powder neutron diffraction
data. Despite fitting to spherically-averaged data, this approach
allows the recovery of the three-dimensional magnetic diffuse scattering
pattern and the spin-pair correlation function. We illustrate the
use of the SPINVERT program with two case studies. First we use simulated
powder data for the canonical Heisenberg kagome model to discuss the
sensitivity of SPINVERT refinement to both pairwise and higher-order
spin correlations. The effect of limited experimental data on the
results is also considered. Second, we re-analyse published experimental
data on the frustrated system Y$_{0.5}$Ca$_{0.5}$BaCo$_4$O$_7$.
The results from SPINVERT refinement indicate similarities between
Y$_{0.5}$Ca$_{0.5}$BaCo$_4$O$_7$ and its parent compound YBaCo$_4$O$_7$,
which were overlooked in previous analysis using powder data.
\end{abstract}

\submitto{\JPCM}

\section{Introduction}

The intense current interest in magnetic materials in which the magnetic
moments (``spins'') do not form a periodic arrangement has been
fuelled by the observation of strongly-correlated states, including
the discovery of emergent magnetic charges in spin-ice materials \cite{Castelnovo_2008,Fennell_2009},
the observation of excitations with fractional quantum number in quantum
spin liquids \cite{Han_2012}, and the presence of chiral order in
otherwise-degenerate states \cite{Machida_2010}. States such as these---lacking
long-range magnetic order, but exhibiting distinctive forms of local
order---have been termed ``cooperative paramagnets'' \cite{Villain_1979}.
The ``gold standard'' experiment for the study of spin correlations
is neutron scattering on a large single crystal sample. Such measurements
provide a direct measurement of the Fourier transform of the three-dimensional
spin correlation function $I(\mathbf{Q})$. In cooperative
paramagnets, neutron scattering experiments often reveal highly-structured
diffuse features in reciprocal space---such as the well-known ``pinch
points'' of spin ice---which can be analysed to characterise the
spin correlations in real-space \cite{Fennell_2009,Isakov_2004}.

Unfortunately, however, large single-crystal samples are not available
for many interesting materials. This is particularly---but far from
exclusively---a problem for newly-synthesised compounds. Consequently,
neutron scattering measurements must often be performed on powder
(polycrystalline) samples. In a powder neutron scattering experiment,
the three-dimensional scattering pattern is collapsed onto a single
axis, $Q$. For magnetic materials which show conventional long-range
spin order, this may not actually present a significant problem, since
the established technique of magnetic Rietveld refinement can (in
favourable cases) provide an unambiguous solution of the magnetic
structure based on only the powder-averaged positions and intensities
of the magnetic Bragg peaks \cite{Wills_2001}. In this way, symmetry
arguments effectively replace the information which is lost as a result
of powder averaging. In cooperative paramagnets, by contrast, the
lack of long-range spin order means that there are no magnetic Bragg
peaks: the powder average of the scattering pattern shows no sharp
features, merely a few diffuse ``humps''. As a consequence, there
has been no obvious analogue of magnetic Rietveld refinement which
can determine the determine the three-dimensional spin correlation
function from powder \emph{diffuse} scattering data. 

In this paper we introduce a new suite of programs, SPINVERT, which
allows determination of the three-dimensional spin correlation function
by fitting powder diffuse scattering data. Our program employs reverse
Monte Carlo refinement \cite{McGreevy_1988,Keen_1996} to fit a large
configuration of spin vectors to experimental powder data. The positions
of spins are fixed at their crystallographic sites throughout the
refinement, while their orientations are refined in order to fit the
data. The RMC algorithm itself is well-established, having previously
been used to study states as diverse as spin glasses \cite{Keen_1996},
ordered magnetic structures \cite{Goodwin_2006}, and flux-line lattices
in superconductors \cite{Laver_2008}. The algorithm is entirely analogous
to a direct Monte Carlo simulation, apart from one important difference:
the function which is minimised during the refinement is not an energy
term defined by a spin Hamiltonian, but rather the sum of squared
residuals which quantifies the level of disagreement between the fit
and experimental data. Thus, RMC is not a technique for modelling
magnetic \emph{interactions} based on a spin Hamiltonian, but rather
a technique for refinement of spin \emph{correlations} based on experimental
data \cite{McGreevy_2001}. 

It is clear even from this brief overview that the RMC approach is
a simple---even na{\" i}ve---one. However, for cooperative paramagnets,
it is also remarkably powerful. In our previous work \cite{Paddison_2012},
we investigated the extent to which three-dimensional information
could be recovered from powder diffuse scattering data in the following
way. First, powder diffuse scattering patterns were simulated for
a number of models of frustrated magnetism. These powder ``data''
were then fitted using the RMC approach. From these refined spin configurations,
we calculated the \emph{three-dimensional} scattering pattern $I(\mathbf{Q})$,
and compared the patterns obtained from RMC with the exact results.
We found that, for each model we considered, almost all the features
of the single-crystal scattering patterns were reproduced by fitting
the powder patterns. Consequently, the spin configurations obtained
by RMC correctly reproduce the three-dimensional spin correlations
of the starting model: they represent one of the degenerate ``solutions''
of the paramagnetic structure.

Why is the RMC approach so successful? To answer this question, we
compare it to alternative ``model-independent'' approaches. Most
such approaches essentially involve fitting a simple form for the
radial spin correlation function to the powder data.
In these methods, the connectivity of the crystal structure is not
considered; instead, either the magnitude of the spin correlations
is fitted for a one-dimensional list of atomic separations (see, \emph{e.g.},
\cite{Park_2003,Schweika_2007}), or the diffuse features are fitted
to a broad peak-shape function to estimate the spin correlation length
(see, \emph{e.g.}, \cite{Granado_2013}). However, the crystal structure
plays no less significant a role in paramagnets than in ordered magnets:
indeed, in a paramagnet the crystal structure completely determines
the symmetry and periodicity of the diffuse scattering pattern \cite{Enjalran_2004}.
Consequently, knowledge of the crystal structure imposes very significant
constraints on the form of both the single-crystal and powder scattering.
Unlike other model-independent approaches---but in common with approaches
using a spin Hamiltonian---RMC refinement uses knowledge of the crystal
structure to constrain the spin correlations.

Our paper is organised as follows. First, we introduce our implementation
of RMC in detail, and summarise the key equations underpinning the
SPINVERT program. Next, we discuss possible ways in which the resulting
spin configurations may be analysed, drawing connections with previous
work using RMC and other techniques. Finally, we present two case
studies intended to illustrate the use of the SPINVERT program. In
the first case study, we fit simulated data for the Heisenberg model
on the kagome lattice \cite{Chalker_1992,Reimers_1993}, and investigate
the effects of limited $Q$-range and statistical errors on the results
which are obtained. In our second case study, we present a new analysis
of existing experimental powder data on Y$_{0.5}$Ca$_{0.5}$BaCo$_4$O$_7$
(from Ref.~\cite{Schweika_2007}) in order to clarify the nature
of the paramagnetic spin correlations in this material. The results
we obtain from SPINVERT analysis indicate some close similarities
between Y$_{0.5}$Ca$_{0.5}$BaCo$_4$O$_7$ and its parent compound
YBaCo$_4$O$_7$, that may have been overlooked in previous
analyses \cite{Schweika_2007}. We conclude with a discussion of the
general advantages and disadvantages of the RMC approach, and some
perspectives for future work. 

\section{Theoretical background}

\subsection*{Reverse Monte Carlo method }

The general RMC method has been described in detail elsewhere \cite{Paddison_2012,Tucker_2007};
here we summarise our specific implementation of magnetic RMC. First,
a supercell of the crystallographic unit cell is generated. The supercell
usually contains several thousand atoms, and periodic boundary conditions
are used to avoid edge effects. A classical spin vector with random
orientation is assigned to each atom, and the goodness-of-fit to experimental
data is calculated:
\begin{equation}
\chi^{2}=W\sum_{Q}\left[\frac{I_{\mathrm{calc}}Q)-I_{\mathrm{expt}}(Q)}{\sigma(Q)}\right]^{2}.\label{eq:chi sq}
\end{equation}
Here, $I(Q)$ is the powder-averaged magnetic scattering
intensity \cite{Blech_1964}, superscript calc and expt
denotes calculated and experimental values, $\sigma(Q)$
is an experimental uncertainty, and $W$ is an empirical weighting
factor. A spin is then chosen at random from the supercell and rotated
by a small amount. This is done by choosing a unit vector $\mathbf{s}$
with random orientation and forming the new spin vector
\begin{equation}
\mathbf{S}_{i}^{\mathrm{new}}=\frac{\mathbf{S}_{i}+\Delta\mathbf{s}}{\left|\mathbf{S}_{i}+\Delta\mathbf{s}\right|},\label{eq:spin_move}
\end{equation}
where $0<\Delta\leq1$ is the maximum spin move length. In practice,
we have found that the final results obtained in RMC fitting are almost
independent of the value of $\Delta$; a value $\Delta=0.2$ is used
by default. After a move is proposed, the change in goodness-of-fit
is calculated, and the proposed move is either accepted or rejected
according to the Metropolis criteria \cite{Metropolis_1949}. This
process is repeated until no further reduction in $\chi^{2}$ is observed.
In this way, spins are iteratively rotated to form a configuration
for which spin correlations are consistent with experimental data.

Throughout this work, we treat spins as having unit length, with the
magnitude of the spin (\emph{i.e.}, the length of the effective magnetic
moment) absorbed into an overall intensity scale factor. When fitting
experimental data, it is almost always necessary to allow the value
of this scale factor $s$ to refine in order to fit the data. The
best-fit value of $s$ can be calculated after each proposed move
by minimising $\chi^{2}$ with respect to $s$, which yields \cite{Proffen_1997}
\begin{equation}
s=\frac{\sum_{Q}\left[I_{\mathrm{calc}}(Q)I_{\mathrm{expt}}(Q)\right]/\left[\sigma(Q)\right]^{2}}{\sum_{Q}\left[I_{\mathrm{calc}}(Q)\right]^{2}/\left[\sigma(Q)\right]^{2}}.\label{eq:scale}
\end{equation}
If the data are placed on an absolute scale, the effective magnetic
moment is then determined by the relation 
\begin{eqnarray}
\mu^2&=&g^2S(S+1)\nonumber \\
&=&s,\label{eq:mag_mon}
\end{eqnarray}
where $g$ is the g-factor. It is possible also to refine a flat or
linear-in-$Q$ background in addition to the scale factor; the relevant
equations are given in Ref.~\cite{Proffen_1997}.

\subsection*{Magnetic neutron scattering intensity}

The magnetic scattering cross-section is variously denoted $I(\mathbf{Q})$,
$S(\mathbf{Q})$ or ${\rm d}\sigma/{\rm d}\Omega$.
Within the quasistatic approximation and for a single type of spin,
it is given as \cite{Squires_1978}:
\begin{eqnarray}
I(\mathbf{Q})&=&C\left[\mu f(Q)\right]^{2}\frac{1}{N}\left|\sum_{i=1}^N\mathbf{S}_i^\perp\exp({\rm i}\mathbf{Q}\cdot\mathbf{r}_i)\right|^{2}.\label{eq:sc_1}
\end{eqnarray}
Here, $f(Q)$ is the magnetic form factor, $\mathbf{Q}$
is the scattering vector (wavevector transfer) and $\mathbf{r}_i$
the position of spin $\mathbf{S}_i$. The lack of divergence of
the magnetic field means that the neutrons only ``see'' the component
of the spin perpendicular to the scattering vector, 
\begin{equation}
\mathbf{S}_i^\perp=\mathbf{S}_i-\left[(\mathbf{Q}\cdot\mathbf{S}_i)\mathbf{Q}\right]/Q^2.
\end{equation}
The proportionality constant $C$ is given by
\begin{eqnarray}
C&=&\left(\frac{\gamma_{\rm n}r_{\rm e}}{2}\right)^2\\
&=&0.07265\,{\mathrm{barn}},
\end{eqnarray}
where $\gamma_{\rm n}$ is neutron magnetic moment in nuclear
magnetons and $r_{\rm e}$ is the classical electron radius.
Multiplying out the squared modulus in Eq.~(\ref{eq:sc_1}) allows
us to write the scattering cross-section in terms of correlations
between pairs of spins,
\begin{equation}
I(\mathbf{Q})=C\left[\mu f(Q)\right]^{2}\left[\frac{2}{3}+\frac{1}{N}\sum_{i,j}\mathbf{S}_i^\perp\cdot\mathbf{S}_j^\perp\cos\left(\mathbf{Q}\cdot\mathbf{r}_{ij}\right)\right],\label{eq:sc2}
\end{equation}
where $\mathbf{r}_{ij}=\mathbf{r}_j-\mathbf{r}_i$. Here, the
double sum over spin pairs excludes the self-correlation terms, for
which $i=j$. These terms give an incoherent contribution to the scattering,
which is the additive factor of $\frac{2}{3}$ in (\ref{eq:sc2}).
In an ideal paramagnet, with zero correlation between different spins,
this incoherent scattering is all that is measured. In a real paramagnet,
however, some degree of spin correlation is still present, often to
temperatures far above the temperature of magnetic ordering. 

\begin{figure}
\begin{centering}
\includegraphics[scale=0.875]{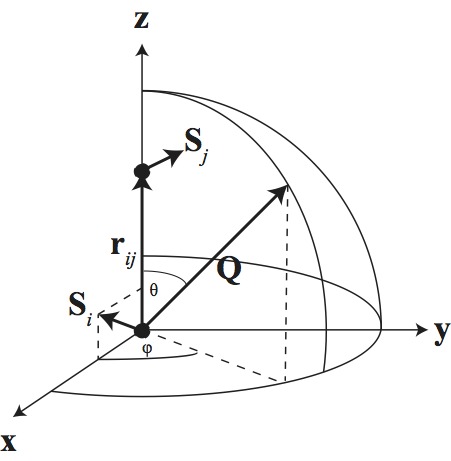}
\par\end{centering}
\caption{\label{fig:spherical coordinates}Local spherical coordinate system
used for calculation of the powder-averaged neutron scattering intensity. }
\end{figure}

In order to calculate the powder scattering cross section, we need
an expression for the spherical average of (\ref{eq:sc2}). An exact
expression was given in the 1960s by Blech and Averbach \cite{Blech_1964},
but appears to have been used only rarely since then. Instead, an
approximate form---applicable only to an isotropic paramagnet---has
often been employed. In this particular case we have 
\begin{equation}
\langle \mathbf{S}_i^\perp\cdot\mathbf{S}_j^\perp\rangle =\frac{2}{3}\langle \mathbf{S}_i\cdot\mathbf{S}_j\rangle,
\end{equation}
so the spin direction is decoupled from $\mathbf{Q}$. The spherical
average over $\theta$ is easily performed by writing 
\begin{equation}
\mathbf{Q}\cdot\mathbf{r}_{ij}=Qr_{ij}\cos\theta.
\end{equation}
 The result is just a sine Fourier transform of the spin correlations,
analogous to the Debye formula or a liquid or a glass \cite{Bertaut_1967}:
\begin{equation}
I_{\rm{isotropic}}(Q)=\frac{2}{3}C\left[\mu f(Q)\right]^2\left(1+\frac{1}{N}\sum_{i,j}\mathbf{S}_i\cdot\mathbf{S}_j\frac{\sin Qr_{ij}}{Qr_{ij}}\right).\label{eq:pw_isotropic}
\end{equation}
It is important to recognise that this expression is exact only for
an isotropic paramagnet in the limit of a large number of spins. It
cannot be used to study materials with magnetic anisotropy (\emph{e.g.},
Ising spins). Since (\ref{eq:pw_isotropic}) is not exact for
a general spin configuration, in SPINVERT we use the exact expression
of Ref.~\cite{Blech_1964}. We sketch their derivation here. Taking
(\ref{eq:sc2}) as the starting point, we define a local coordinate
system for each pair of spins, $i,j$. The local $\mathbf{z}$ axis
is directed along the vector separating the pair of spins; the local
$\mathbf{x}$ axis is perpendicular to $\mathbf{z}$ and lies the
plane of $\mathbf{S}_i$; and $\mathbf{y}$ is the remaining vector
in a right-handed set. This gives for the coordinate axes: $\mathbf{z}=\mathbf{r}_{ij}/r_{ij}$,
$\mathbf{x}=\left[\mathbf{S}_i-\left(\mathbf{S}_i\cdot\mathbf{z}\right)\mathbf{z}\right]/\left|\mathbf{S}_i-\left(\mathbf{S}_i\cdot\mathbf{z}\right)\mathbf{z}\right|$,
and $\mathbf{y}=\mathbf{z}\times\mathbf{x}$. With this local coordinate
system, we apply the identity 
\begin{equation}
\mathbf{S}_i^\perp\cdot\mathbf{S}_j^\perp=\sum_{\alpha,\beta}\left(\delta^{\alpha\beta}-\frac{Q^{\alpha}Q^{\beta}}{Q^{2}}\right)S_i^{\alpha}S_j^{\beta}\label{eq:perp2}
\end{equation}
where, from Fig.~\ref{fig:spherical coordinates}, the components
$\alpha,\beta\in[x,y,z]$ of $\mathbf{Q}$ are given in
spherical coordinates by
\begin{equation}
\mathbf{Q}=[Q\sin\theta\cos\varphi,Q\sin\theta\sin\varphi,Q\cos\theta].\label{eq:q_cmpts}
\end{equation}
Substituting (\ref{eq:q_cmpts}) into (\ref{eq:perp2}), and the result
into (\ref{eq:sc2}), gives an expression for $I\left(\mathbf{Q}\right)=I\left(Q,\theta,\varphi\right)$.
The integrals over $\theta$ and $\varphi$ can then be done by hand,
which gives the final result 
\begin{equation}
I\left(Q\right)=C\left[\mu f(Q)\right]^2\left\{\frac{2}{3}+\frac{1}{N}\sum_{i,j}\left[A_{ij}\frac{\sin Qr_{ij}}{Qr_{ij}}+B_{ij}\left(\frac{\sin Qr_{ij}}{\left(Qr_{ij}\right)^{3}}-\frac{\cos Qr_{ij}}{\left(Qr_{ij}\right)^{2}}\right)\right]\right\}. \label{eq:powder}
\end{equation}
in which
\begin{eqnarray}
A_{ij}&=&S_i^xS_j^x,\\
B_{ij}&=&2S_i^zS_j^z-S_i^xS_j^x.
\end{eqnarray}
The sum is taken over all pairs of spins which are separated by radial
distances $0<r_{ij}\leq r_{\rm {max}}$; when periodic boundary
conditions are imposed, the maximum radial distance $r_{\rm {max}}$
is given by half the length of the shortest side of the RMC supercell
(``box''). 

We end this section by anticipating a common criticism of the RMC
approach. It is sometimes argued that the use of a large supercell
means that the number of parameters $p$ to be estimated greatly exceeds
the number of data points $d$. The number of parameters is often
counted as $p=2N$ for $N$ spins (since each spin has two rotational
degrees of freedom), which indeed is usually greater than $n$. It
is implicit here that the number of degrees of freedom of the RMC
fit is calculated using the standard expression $\nu=n-p$. However,
this is only applicable when the function which is fitted to the data
is linear in the parameters. Since the magnetic scattering cross-section
depends on the dot product of pairs of spins, this is not the
case, and the effective number of parameters is not well-defined.
Moreover, the quantity of interest from Monte Carlo simulations---both
direct and reverse---is also not the orientations of individual spins,
but rather the correlations between spin-pairs. Consequently,
the effectiveness of RMC is not well assessed by such arguments, and
is better judged by its performance in real-world applications.

\section{Performing SPINVERT refinements: some guidelines}

In order to perform a refinement using SPINVERT only a few essential
pieces of information are required. First, we need four pieces of
experimental information about the system being studied: 
\begin{enumerate}
\item{
\emph{Experimental powder diffuse scattering data.} The input data
should contain only \emph{magnetic} diffuse scattering. Ideally the
data will have be obtained using neutron polarisation analysis to
isolate the magnetic signal \cite{Scharpf_1993}. However, data for
which the magnetic signal has been obtained by subtraction of a very
high-temperature dataset from a lower-temperature dataset can also
used. We emphasise that the quality of the results ultimately obtained
is entirely dictated by the quality of the data; some guidelines in
this respect are given in Section~\ref{sec:Kagome-Heisenberg-magnet}.
}
\item{
\emph{Crystallographic details: lattice parameters and fractional
coordinates of atoms in the crystallographic unit cell.} The SPINVERT
program requires that the specified unit cell has orthogonal axes.
For hexagonal or rhombohedral systems it is necessary to define a
larger orthogonal unit cell for use in SPINVERT. It is further assumed
that all atoms in the unit cell have identical magnetic properties
(magnetic moments and magnetic form factors), though all atoms need
not be crystallographically equivalent. The atom positions are not
all required to be occupied by a spin, so it is possible to refine
data for systems with substitutional (but not positional) disorder.
}
\item{
\emph{Magnetic form factor.} An accurate parameterisation of the magnetic
form factor is likely to prove important in SPINVERT refinement, probably
to a greater extent than in conventional Rietveld refinement \cite{Rotter_2009}.
If the magnetic form factor is not well approximated by tabulated
functions, it may be worth performing an accurate measurement to determine
it before proceeding.
}
\item{
\emph{Single-ion anisotropy.} Spin anisotropy may be fixed by giving
the direction of the easy axis for each atom in the unit cell (for
Ising spins), or the direction perpendicular to the easy plane (for
XY spins). Even if Heisenberg spins are specified, it is still possible
for the refinement to produce anisotropic spin configurations, should
this be required to fit the data. However, there are almost certain
to be many more ways of fitting the data in which spins do \emph{not}
point along their anisotropy axes, so SPINVERT is likely to underestimate
the extent of magnetic anisotropy \cite{Paddison_2012}. Therefore,
if other measurements indicate that spins behave as essentially pure
Ising or XY variables, we suggest constraining the spin anisotropy
accordingly.
}
\end{enumerate}
There are also four main options concerning the SPINVERT refinement
itself:
\begin{enumerate}
\item{
\emph{Weight.} The weight, $W$ in (\ref{eq:chi sq}), is equivalent
to inverse temperature in a direct MC simulation, and determines the
proportion of ``bad'' moves that are accepted despite leading to
an increase in $\chi^{2}$---that is, how closely the data as a whole
are fitted. Its value can be freely specified by the user. In practice,
the value of $W$ should be chosen such that the fit reproduces the
real features of the data without over-fitting to the experimental
errors. A good starting point is often to choose $W$ such that between
25\% and 75\% of the ``bad'' moves are accepted when the refinement
has converged.
}
\item{
\emph{Box size.} The size of the spin configuration, given by the
number of unit cells along each Cartesian direction which make up
the RMC supercell. The maximum correlation length which can be modelled
is given by one half the length the shortest side of the box, so a
larger box is needed to reproduce sharper features in $I\left(Q\right)$.
However, once a sufficiently large box has been identified, there
is nothing to be gained by increasing the box size further---doing
so merely gives the refinement additional degrees of freedom to vary
the long-range spin correlations, which it may do by fitting to high-frequency
components (\emph{e.g.}, noise) in the experimental data.
}
\item{
\emph{Maximum simulation time.} In general, refinements should be
run until no reduction is seen in $\chi^{2}$. The number of moves
required depends on the data, the system and the refinement parameters.
In practice, $\sim$10$^3$ moves per spin is often sufficient.}
\item{
\emph{Intensity scale and background corrections.} When refining experimental
data it is almost always necessary to treat the overall intensity
scale as a fitting parameter. This is the case when the data have
not been placed on an absolute intensity scale, and/or the magnitude
of the magnetic moment is not known precisely. In some cases, the
minimum in $\chi^2$ can be quite flat as a function of the scale.
Then the uncertainty in the scale may be the main source of error
in quantities derived from the refinement (\emph{e.g.}, spin correlation
length). The uncertainty in derived quantities may be determined,
for an assumed uncertainty in the scale, by performing separate refinements
with the scale factor fixed at (say) $\pm5\%$ of its refined value.
It is also possible to refine a flat or linear-in-$Q$ background
term; however, this should not be done unless necessary, since it
introduces an extra parameter into the fit.
}
\end{enumerate}

The key output from SPINVERT is a spin configuration---a file containing
a list of spin vectors, with their corresponding positions in the
RMC supercell. An individual spin configuration represents a snapshot
picture of a possible arrangement of spins which is consistent with
experimental data. In this sense, a spin configuration represents
a magnetic structure ``solution''. A direct consequence of the absence
of long-range magnetic order is that this magnetic structure solution
cannot be unique: running the reverse Monte Carlo program again will
produce a different spin configuration, which may be thought of as
a second snapshot taken a long time after the first. In fact, it is
generally important to run RMC simulations several times and average
calculated quantities, so that a degree of statistical confidence
can be obtained in the results. In the following section, we consider
how the spin configurations obtained from RMC can be mined for information
about the system properties.

\section{Analysis of paramagnetic spin configurations }

Once RMC refinements have been performed, the next---and most interesting---stage
of the analysis can proceed: unravelling the actual physics which
is contained in the spin configurations. Since paramagnetic spin configurations
are not unique or periodic, there is no way to describe them that
is at once comprehensive and general while using few parameters. This
point is equally applicable to disordered spin configurations obtained
from other techniques (such as direct Monte Carlo simulation), and
reflects a general problem of how to characterise disordered states
\cite{Cliffe_2010,Billinge_2010}. To obtain the fullest understanding possible,
it is often useful to calculate quantities in both real and reciprocal
space: the real-space picture identifies the local spin correlations,
whereas the reciprocal-space picture identifies periodicities and
long-range components to the correlations \cite{Bramwell_2011}. To
facilitate this kind of analysis, the associated program SPINCORREL
plots real-space spin correlation functions, and the program SPINDIFF
calculates the three-dimensional magnetic scattering pattern $I(\mathbf{Q})$.

\subsection*{Real-space: spin correlation functions}

It is tempting to try examining directly the orientations of spins
within the configuration, but the lack of periodicity means that meaningful
patterns are often hard to discern. Instead, it is better to plot
spin-pair correlation functions, which provide a measure of how some
property of spin-pairs varies as a function of the separation of the
pair of spins. In general,
\begin{equation}
\langle f(\mathbf{S}(\mathbf{0}),\mathbf{S}(\mathbf{r}))\rangle =\frac{1}{N}\sum_i^Nf(\mathbf{S}_i,\mathbf{S}_j),\label{eq:correl_fn}
\end{equation}
where $f(\mathbf{S}_i,\mathbf{S}_j)$ is some function
of a central spin $\mathbf{S}_i$ and a spin $\mathbf{S}_j$ at
vector $\mathbf{r}$ away from it, and an average is performed over
all spins in the configuration as centres. The powder diffraction
pattern is essentially a direct measurement of the Fourier transform
of the radial spin correlation function $\langle \mathbf{S}(0)\cdot\mathbf{S}(r)\rangle =\langle \cos\varphi\rangle$,
which measures the average scalar product of pairs of spins separated
by distance $r$. For a spin configuration, this average is performed
by taking each spin $\mathbf{S}_i$ in turn, and then averaging
the spin correlations over its $Z_{ij}$ neighbours, $\mathbf{S}_j$,
which coordinate it at distance $r$:
\begin{equation}
\langle\mathbf{S}(0)\cdot\mathbf{S}(r)\rangle =\frac{1}{n(r)}\sum_{i}^{N}\sum_{j}^{Z_{ij}(r)}\mathbf{S}_i\cdot\mathbf{S}_j,\label{eq:radial_scf}
\end{equation}
where $n(r)$ is given by 
\begin{equation}
n(r)=\sum_i^NZ_{ij}(r),\label{eq:num_r}
\end{equation}
with $Z_{ij}(r)$ the number of spins $j$ which coordinate
a central spin $i$ at distance $r$.

There are many other correlation functions which have been invented
to demonstrate various aspects of structures; some examples are considered
in the case studies which follow. In structures involving triangular
motifs it may be useful to plot the correlation function of the vector
or scalar chirality, defined in Ref.~\cite{Reimers_1993}, or the
spin nematic correlation function [Section~\ref{sec:Kagome-Heisenberg-magnet}].
In polarised neutron scattering experiments on single-crystal samples
it is possible to separate the different Cartesian components of the
scattering function with different crystal orientations \cite{Fennell_2009};
the spin-flip and non-spin-flip scattering cross-sections which are
measured are also accessible from SPINVERT refinements. In Section~\ref{sec:Kagome-Heisenberg-magnet}
we will consider the \emph{three-dimensional} spin correlation function,
the real-space analogue of the three-dimensional scattering function
$I(\mathbf{Q})$. Finally, we note that SPINVERT refinements
provide access not only to the configurational average of correlation
functions, but also to their distributions. For example, $\langle \cos\psi\rangle =-0.5$
may well indicate $120^{\circ}$ correlations, but could also result
from a bimodal distribution of correlations centred at $90^{\circ}$
and $180^{\circ}$ .

\subsection*{Reciprocal space: magnetic scattering functions}

The three-dimensional magnetic scattering pattern $I(\mathbf{Q})$
can reveal important properties of frustrated magnets that are not
always so clear in real-space---for example, the power-law correlations
in many frustrated magnets appear as ``pinch-points'' in $I(\mathbf{Q})$.
In principle, calculating $I(\mathbf{Q})$ from a spin
configuration is straightforward using (\ref{eq:sc_1}) or (\ref{eq:sc2}).
While both these equations are mathematically equivalent, it is much
faster to compute $I(\mathbf{Q})$ using (\ref{eq:sc_1})
(which requires $\mathcal{O}(N)$ operations) rather than
(\ref{eq:sc2}) (which requires $\mathcal{O}(N^2)$ operations).
However, choosing to calculate (\ref{eq:sc_1})  leads to some complications
in practice. First, it is not possible to allow for periodic boundary
conditions. This is obviously problematic for refinements which were
performed with periodic boundary conditions applied. Second, when
the size of the RMC box is large compared to the spin correlation
length of the system, the calculated scattering pattern appears very
noisy. This seems counterintuitive---since using a larger supercell
will reduce the error in $\langle\mathbf{S}(\mathbf{0})\cdot\mathbf{S}(\mathbf{r})\rangle $---but
the problem is that as the supercell size is increased, so too does
the number of inter-atomic vectors which exceed the spin correlation
length. The Fourier transform of these long-wavelength components
appears as high-frequency noise in the calculated scattering pattern.
In a real experiment, by contrast, this effect is averaged out by
the instrumental $\mathbf{Q}$-resolution \cite{Butler_1992}.

In fact, both these observations suggest the same thing: that it is
necessary to restrict the maximum range of the interatomic vectors
considered in (\ref{eq:sc_1}). When periodic boundary conditions
are applied, the maximum interatomic vector should not exceed half
the box size in each direction. A smaller cutoff may be chosen to
correspond to the length-scale at which spin correlations become negligible.
The question of how to impose such a cutoff was originally addressed
in \cite{Butler_1992}, where it was pointed out that the noise in
calculated scattering patterns could be reduced by dividing the RMC
supercell into a set of smaller regions called ``sub-boxes'' or
``lots'', and averaging the scattering intensity over all sub-boxes.
In effect, this procedure applies a box filter in real space. The
calculation proceeds by calculating the three equations in sequence:
\begin{equation}
\mathbf{A}_{\mathbf{R}}(\mathbf{Q})=\sum_{a=1}^{N_{a}}\mathbf{S}_{\mathbf{R},a}^\perp\exp(\mathrm{i}\mathbf{Q}\cdot\mathbf{r}_a),\label{eq:A_Q}
\end{equation}
where $\mathbf{r}_a$ is the position of spin $\mathbf{S}_{a,\mathbf{R}}$
within a crystallographic unit cell, $\mathbf{R}$ is a lattice vector
giving the origin of the sub-box within the RMC supercell, and the sum runs over the $N_a$
atoms in the unit cell; 
\begin{equation}
\mathbf{M}_{\mathbf{R}}(\mathbf{Q})=\sum_{\mathbf{R^{\prime}=}\mathbf{R}}^{\mathbf{R+L}}\mathbf{A}_{\mathbf{R^{\prime}}}(\mathbf{Q})\exp(\mathrm{i}\mathbf{Q}\cdot\mathbf{R^{\prime}}),\label{eq:M_Q}
\end{equation}
where $\mathbf{L}$ is the sub-box size; and finally a sum over all
sub-boxes:
\begin{equation}
I(\mathbf{Q})=\sum_{\mathbf{R}}\left|\mathbf{M}_{\mathbf{R}}(\mathrm{\mathbf{Q}})\right|^2.\label{eq:I_Q}
\end{equation}
Here, we have used the lattice periodicity to write $\mathbf{r}_i=\mathbf{r}_a+\mathbf{R}$. 

The method of sub-boxes greatly reduces the level of noise, and produces
calculated scattering patterns in good agreement with experimental
data. However, the original method of \cite{Butler_1992} also has
some disadvantages. In particular, the origin of each sub-box was
chosen to be a random lattice vector of the supercell with position
$\mathbf{R}$, so a quantitatively different answer is obtained each
time the calculation is performed. This problem can be avoided by
using every lattice vector $\mathbf{R}$ in the supercell as the origin
of a sub-box. However, a straightforward implementation of such an
algorithm is very slow. In SPINDIFF, we accelerate the calculation
using a fast blurring algorithm often employed in gaming and computer
graphics. This takes advantage of the fact that adjacent lots contain
the same set of spins, apart from at the lot edges, so that in one
dimension we can write 
\begin{equation}
\mathbf{M}_{x+1}(\mathbf{Q})=\mathbf{M}_x(\mathbf{Q})+\mathbf{A}_{x+L+1}-\mathbf{A}_{x}.\label{eq:motion_blur}
\end{equation}
The key point is that applying (\ref{eq:motion_blur}) along each
of the Cartesian directions in sequence generates (\ref{eq:M_Q})
without redundant calculations.

While the method of sub-boxes can improve the appearance of the calculated
$I(\mathbf{Q})$, it remains important to have sufficient
statistical averaging. Calculated patterns usually appear noisy if
they are obtained using only a few thousand spins; to obtain good
statistics, it is usually necessary to include $\sim$10$^5$ spins
in the calculation. The easiest way of achieving this is to average
the $I(\mathbf{Q})$ over many individual spin configurations. 

As was the case during the SPINVERT refinement, we make some assumptions
when calculating $I(\mathbf{Q})$:
\begin{itemize}
\item{
The program calculates only the magnetic scattering; it does not calculate
any nuclear scattering.
}
\item{
The positions of the spins are fixed at their crystallographic positions. 
\item There is no magnetoelastic coupling.
}
\item{
All the spins have identical magnetic properties (magnetic form factors
and magnetic moments).
}
\end{itemize}
Provided these assumptions are satisfied, the program SPINDIFF can
be used to calculate the magnetic diffuse scattering for any spin
configuration, including those obtained from direct Monte Carlo simulations,
spin density-functional theory, and other computational approaches.

\section{Kagome Heisenberg magnet \label{sec:Kagome-Heisenberg-magnet}}

As a first case study we consider the Heisenberg nearest-neighbour
antiferromagnet on the kagome lattice. This canonical frustrated model
system was studied in detail in the early 1990s, and does not show
long-range magnetic order at finite temperature. At low temperature,
spins on each triangle form a coplanar $120^{\circ}$ arrangement,
and there is a macroscopic degeneracy of ways in which these coplanar
units can be arranged \cite{Chubukov_1992,Chalker_1992}. Several
real materials seem to realise aspects of this behaviour with either
classical or quantum spins---for example, hydronium jarosite \cite{Wills_1996,Fak_2008}
and herbertsmithite \cite{Han_2012,Vries_2009}---although in practice
there is usually some subtle feature which breaks the degeneracy of
ground states.

In order to demonstrate the capabilities of SPINVERT, we first simulate
powder diffuse scattering ``data'' for this model. We consider the
classical spin Hamiltonian with nearest-neighbour interactions,
\begin{equation}
H=-\frac{1}{2}J\sum_{\langle i,j\rangle }\mathbf{S}_i\cdot\mathbf{S}_j,\label{eq:nn_hamiltonian}
\end{equation}
at a low temperature $T/J=0.01$. Direct Monte Carlo simulations for
this Hamiltonian were performed for spin configurations of size $3600$
spins (comprising ten stacked kagome slabs each of size 360 spins)
using periodic boundary conditions. The simulation time was $10^{5}$
proposed MC moves per spin, and 16 independent simulations were performed
in order to obtain good statistics for the scattering calculations.
The powder diffuse scattering pattern was then calculated from the
MC spin configurations using (\ref{eq:powder}), with an arbitrary
magnetic form factor (Ho$^{3+}$) applied. In order to
facilitate comparison between powder and single-crystal diffuse scattering
patterns, we plot powder scattering patterns as a function of the
dimensionless parameter $Q^\prime=aQ/2\pi$, and label high-symmetry
points $(hkl)$ in the single-crystal patterns. Powder
data were simulated in the range $0\,<Q^\prime\leq5$, and binned
at an interval $\Delta Q^\prime=0.02$. 

\begin{figure}
\begin{centering}
\includegraphics[scale=0.875]{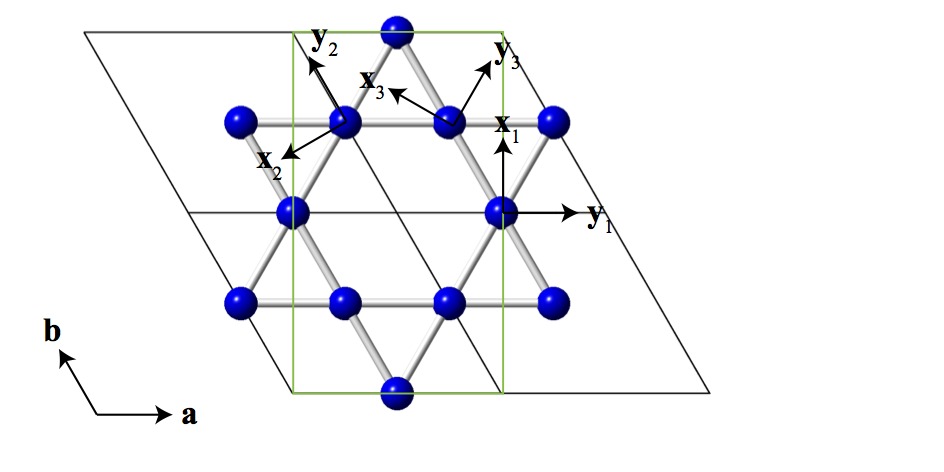}
\par\end{centering}

\caption{\label{fig:kagome_structure}Kagome lattice. The hexagonal unit cell
is shown in black, the orthorhombic cell used for RMC refinements
in green, and local basis vectors $\mathbf{x}$ used to determine
are shown as red arrows. The local $\mathbf{x}$ axis for a given
atom points towards the centre of one of the triangles to which the
atom belongs, and the local $\mathbf{y}$ axis is a perpendicular
vector. Note that due to the point symmetry $\mathbf{x}$ and $-\mathbf{x}$
are equivalent directions, as are $\mathbf{y}$ and $-\mathbf{y}$.}
\end{figure}

\begin{figure}
\begin{centering}
\includegraphics[scale=0.875]{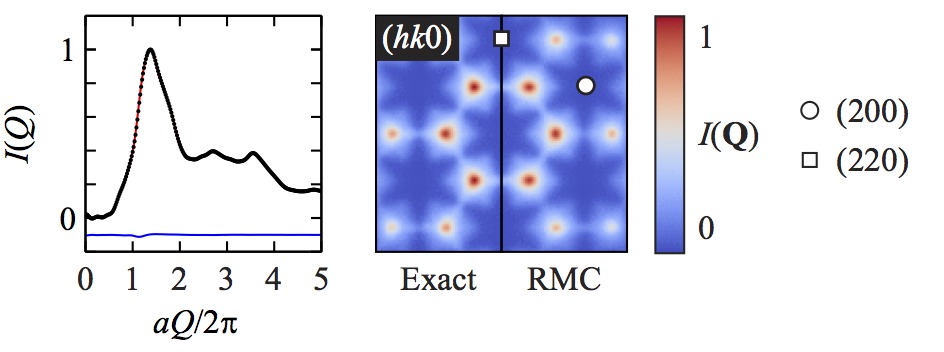}
\par\end{centering}

\caption{\label{fig:kagome_fit}Left: SPINVERT fit to powder data for the antiferromagnetic
Heisenberg model on the kagome lattice. Input data are shown as black
circles, fit as a a red line, and difference (data--fit) as a blue
line. Right: Single-crystal scattering in the $(hk0)$
reciprocal-space plane. The panel labelled ``Exact'' is calculated
from the model used to generate the input data (left image); the panel
labelled ``RMC'' is calculated from the SPINVERT refinement of these
simulated powder data, as discussed in the text. Both panels are on
the same intensity scale.}
\end{figure}

Having obtained these simulated ``data'' we use them as input data
for SPINVERT refinements. Since SPINVERT works with orthogonal axes,
it is necessary to convert the hexagonal unit cell with dimensions
to an orthorhombic cell containing twice as many atoms; the transformation
matrix for this (frequently-encountered) case is given by
\begin{equation*}
\left[\begin{array}{c}
\mathbf{a^{\prime}}\\
\mathbf{b^{\prime}}\\
\mathbf{c^{\prime}}
\end{array}\right]=\left[\begin{array}{ccc}
1 & 0 & 0\\
1 & 2 & 0\\
0 & 0 & 1
\end{array}\right]\left[\begin{array}{c}
\mathbf{a}\\
\mathbf{b}\\
\mathbf{c}
\end{array}\right].
\end{equation*}
During the following refinements we set the overall intensity scale
factor as a parameter to be fitted. Although the absolute scale is
already known for the simulated data considered here, this is very
unlikely to be the case with experimental data, so we follow the general
procedure of allowing a scale factor to refine. We performed SPINVERT
refinements for 300 moves per spin, after which time no significant
improvement was observed in the value of $\chi^{2}$. The fit-to-data
obtained using RMC is shown in Fig.~\ref{fig:kagome_fit}(a), showing
essentially perfect agreement with the data. In fact, this is usually
the case for RMC refinements; indeed, if a refinement fails to fit
a significant feature in a high-quality dataset, this is usually an
indication that something is wrong with the starting model or the
refinement parameters.

Having obtained spin configurations in agreement with experiment,
we consider how well these fitted spin configurations reproduce the
known features of the kagome Heisenberg model. Looking first at reciprocal
space, we calculate the single-crystal scattering pattern $I(\mathbf{Q})$
from the SPINVERT refinement and compare it with the corresponding
exact pattern obtained from the direct MC simulations [Fig.~\ref{fig:kagome_fit}(b)].
Both patterns were calculated using SPINDIFF. Excellent agreement
is obtained between the exact result and the SPINVERT prediction,
with only a slight blurring of the pattern---corresponding to disordering
of the spins---notable in the SPINVERT pattern. We note that this
ability to recover the three-dimensional scattering pattern from spherically-averaged
data is not limited to this particular model system. In our previous
work \cite{Paddison_2012}, we considered seven separate systems with
different crystal structures and magnetic interactions, and found
that the three-dimensional scattering pattern was recoverable from
powder data in each case.

\begin{figure}
\begin{centering}
\includegraphics[scale=0.875]{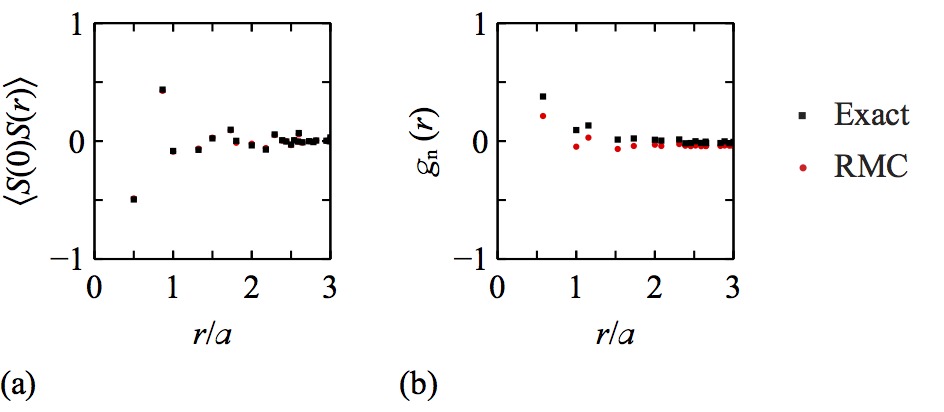}
\par\end{centering}

\caption{\label{fig:kagome_scf_ncf}(a) Radial spin correlation function $\langle\mathbf{S}(0)\cdot\mathbf{S}(r)\rangle$. Exact values (from direct Monte Carlo simulations) are shown as
black squares, and values obtained from the SPINVERT fit to powder
data are shown as red circles. The SPINVERT values are near-identical
to the exact values. (b) Nematic spin correlation function $g_{\rm n}(r)$
defined in the text. Exact values are shown as black squares, and
values obtained from the SPINVERT fit are shown as red circles. The
SPINVERT fit correctly determines the trend in nematic correlations,
but underestimates the magnitude by $\sim$50\%.}
\end{figure}

We now ask how the spin structures obtained by SPINVERT can be characterised
in real space. In Fig.~\ref{fig:kagome_scf_ncf}(a) we plot the radial
spin correlation function $\langle\mathbf{S}(0)\cdot\mathbf{S}(r)\rangle$
for the SPINVERT refinements, and compare it with the exact result.
The SPINVERT correlation function is nearly identical to the exact
result, typically underestimating the magnitude of the correlations
by a few percent. Thus it appears that powder data are very effective
at determining the radial spin correlations. While this result is
perhaps unsurprising, it is not necessarily the case, since $I(Q)$
and $\langle\mathbf{S}(0)\cdot\mathbf{S}(r)\rangle$
only contain equivalent information in the limit of large $Q_{\rm{max}}$. 

Following on from this, a natural question is whether the RMC refinements
might afford some sensitivity to higher-order spin correlations. Such
correlations are indeed present in the kagome Heisenberg AFM model.
At low temperatures $(T/J\lesssim0.02)$, short-range spin
\emph{nematic} order develops, so that adjacent triangles of spins
tend to lie in a common plane; the mechanism may involve ``order
by disorder'' \cite{Chalker_1992,Reimers_1993}. The ``standard''
spin correlation function is insensitive to nematic order, so, following
Ref.~\cite{Chalker_1992}, we define a nematic correlation function
which is equal to 1 for a coplanar state:
\begin{equation}
g_{\mathrm{n}}\left(r\right)=\frac{3}{2}\left\langle \left(\mathbf{n}\left(0\right)\cdot\mathbf{n}\left(r\right)\right)^{2}\right\rangle -\frac{1}{2},
\end{equation}
where
\begin{equation}
\mathbf{n}=\frac{2}{3\sqrt{3}}(\mathbf{S}_1\times\mathbf{S}_2+\mathbf{S}_2\times\mathbf{S}_3+\mathbf{S}_3\times\mathbf{S}_1)
\end{equation}
is the vector normal to the plane formed by the three spins of a triangle.
The comparison of the SPINVERT $g_{\rm n}(r)$ with
the exact result is shown in Fig.~\ref{fig:kagome_scf_ncf}(b). It
is apparent that SPINVERT consistently underestimates the value of
$g_{\rm n}(r)$ by around 50\%. Thus it seems that
SPINVERT is somewhat less sensitive to higher-order spin correlations
than to spin-pair correlations. Conversely, since the experimental
data are only directly sensitive to spin-pair correlations, it is
perhaps surprising that RMC should show any sensitivity to higher-order
correlations at all.

\begin{figure}
\begin{centering}
\includegraphics[scale=0.875]{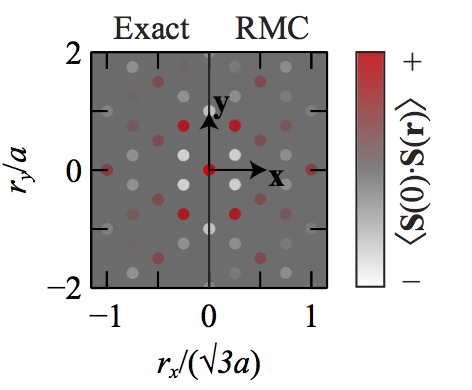}
\par\end{centering}

\caption{\label{fig:kagome_3d_scf}Two-dimensional spin correlation function
$\langle\mathbf{S}(\mathbf{0})\cdot\mathbf{S}(\mathbf{r})\rangle$
for the Heisenberg kagome antiferromagnet. The left panel shows exact
values, and the right panel shows values obtained from SPINVERT refinement
of powder data. The $x$ and $y$ axes are the local axes
given in Fig.~\ref{fig:kagome_structure}.}
\end{figure}

Finally, we consider the directional dependence of the real-space
spin correlations. The correlation functions we have plotted so far
are radial averages, which take no account of the geometry of the
kagome lattice. However, this must obscure a great deal of information---after
all, it is the geometry of the lattice which leads to frustration.
In order to represent the directional dependence of the correlations,
we consider the two-dimensional spin correlation function $\langle\mathbf{S}(\mathbf{0})\cdot\mathbf{S}(\mathbf{r})\rangle$.
This function is seldom shown in the literature (although its Fourier
transform $I(\mathbf{Q})$ is frequently given), but it
can reveal surprisingly simple patterns in the paramagnetic structure
\cite{Paddison_2013}. For a Bravais lattice, $\langle\mathbf{S}(\mathbf{0})\cdot\mathbf{S}(\mathbf{r})\rangle$
is straightforward to calculate: one simply calculates the dot product
of a central spin with the spin at displacement $\mathbf{r}$ away,
and averages over each spin as a centre. For a non-Bravais lattice,
such as kagome, the calculation is slightly more involved, because
different basis atoms can be related to each other by rotational as
well as translational symmetries. In such cases, it is necessary to
rotate our coordinate axes when moving between basis atoms such that
each atom is in an equivalent environment. For the kagome lattice,
with a three-atom basis, a set of suitable local axes are shown in
Fig~\ref{fig:kagome_structure}. The directional dependence of the
spin correlations is shown in Fig.~(\ref{fig:kagome_3d_scf}). The
``Exact'' spin correlations for the starting model [Fig.~(\ref{fig:kagome_3d_scf}),
left panel] show a non-trivial dependence on both distance and angle,
yet the result from SPINVERT refinement reproduces all the main features
of this dependence [Fig.~(\ref{fig:kagome_3d_scf}), left panel].

An obvious question which remains is: how well do these results transfer
to actual experimental data? Compared with the simulated data used
above, real data are likely to have a somewhat more limited $Q$-range,
and to contain a certain amount of statistical noise. We will now
consider the effects of these limitations, and make some suggestions
as to the kind of data which are suitable for SPINVERT refinement.

\begin{figure}
\begin{centering}
\includegraphics[scale=0.875]{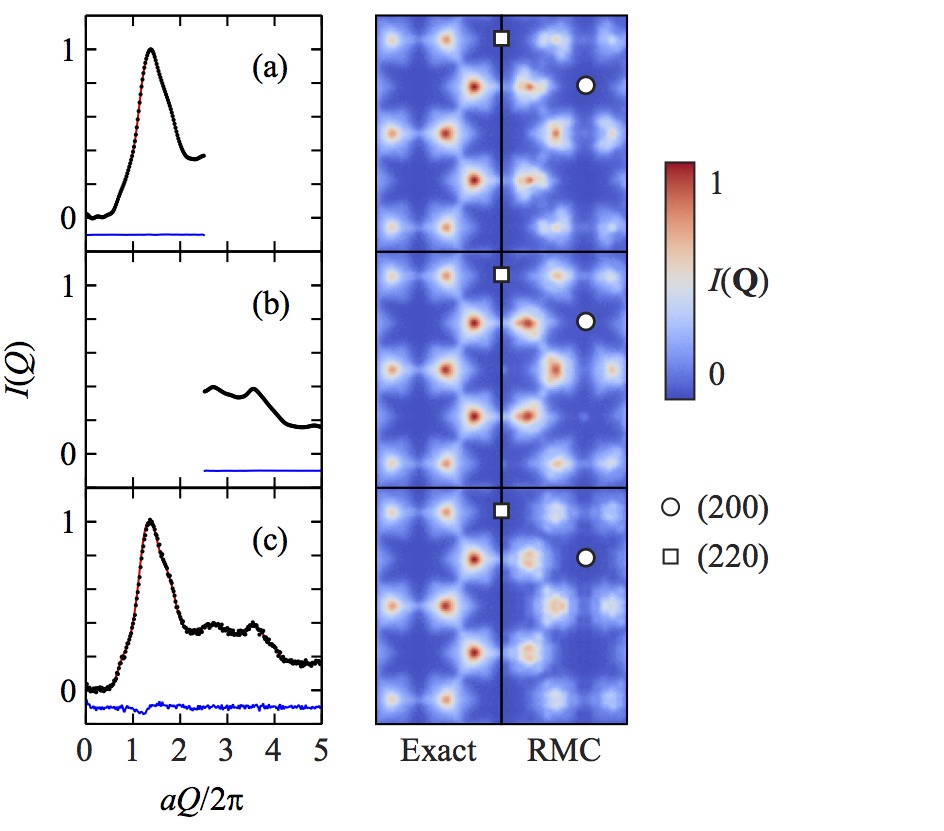}
\par\end{centering}

\caption{\label{fig:kagome_q_noise}(a) Fit to powder data and predicted single-crystal
scattering pattern for input powder data with limited $Q_{\mathrm{max}}^{\prime}=2.5$.
Input data are shown as black circles, fit as a a red line, and difference
(data--fit) as a blue line. (b) Fit to powder data and predicted single-crystal
scattering pattern for input powder data with $Q_{\mathrm{min}}^{\prime}=2.5$.
(c) Fit to powder data and predicted single-crystal scattering pattern
for input powder data containing statistical noise, as described in
the text.}
\end{figure}

We investigate the effect of limited $Q$-range on the predicted single-crystal
scattering patterns by identifying two extreme scenarios. First, we
attempt a refinement omitting all the data except the main diffuse
peak, by setting $Q_{\mathrm{max}}^\prime=2.5$ [Fig.~\ref{fig:kagome_q_noise}(a)].
Then we consider the inverse situation, omitting all the data below
$Q_{\mathrm{min}}^{\prime}=2.5$ but keeping the subsequent peaks
[Fig.~\ref{fig:kagome_q_noise}(b)]. As anticipated, in both
cases the reconstruction of the single-crystal scattering is of lower
quality than for the full refinement. What is surprising is that both
refinements nevertheless reproduce the main features of the diffuse
scattering. Moreover, the overall quality of both reconstructions
is similar, even though the input data do not overlap in $Q$. This
behaviour can be understood by recognising that the diffuse scattering
pattern is periodic in reciprocal space, and this periodicity is dictated
by the crystal structure \cite{Enjalran_2004}. Consequently, the
powder-averaged scattering from subsequent Brillouin zones contains
information about the first Brillouin zone, and \emph{vice versa}.
In effect, every measured value of $Q$ represents an average over
a different selection of points in the first Brillouin zone, so that
increasing $Q_{\mathrm{max}}$ places increasingly stronger restrictions
on $I(\mathbf{Q})$. 

Does this mean it is best to measure data to the highest possible
$Q_{\mathrm{max}}$? In practice, the answer is probably only a qualified
``yes''. An exception is that the behaviour of the single-crystal
scattering in the limit $\mathbf{Q}\rightarrow\mathbf{0}$ is uniquely
determined by the powder scattering as $Q\rightarrow0$. This limit
is generally less experimentally accessible for larger values of $Q_{\mathrm{max}}$.
As a result, if there is insufficient information in the data as a
whole to constrain the low-$Q$ behaviour, there is a tendency for
SPINVERT refinement to refine inaccurate values for $I(Q\rightarrow0)$.
An example of this behaviour can be seen in Fig.~\ref{fig:kagome_q_noise}b.
Another consideration is that, as the scattering becomes weaker at
higher $Q$, the experimental errors become proportionately larger,
and fitting to errors in the high-$Q$ data may produce noise even
at low-$Q$ in the predicted single-crystal patterns. We may therefore
ask how low the statistical errors need to be. To test this, we added
a small amount of statistical noise to our simulated data and repeated
the SPINVERT refinement. The noise was simulated by choosing random
numbers from a Gaussian distribution with standard deviation $\sigma=0.01$
(corresponding to an error of 1\% in the most intense data point).
With these (relatively small) errors, we find that the single crystal
pattern predicted by fitting the powder data is noticeably reduced
in quality [Fig.~\ref{fig:kagome_q_noise}(c)], although the
main features are still recovered.

We therefore draw two main conclusions regarding the data quality
which is desirable for SPINVERT refinement. First, the differences
between Fig.~\ref{fig:kagome_fit} and Fig.~\ref{fig:kagome_q_noise}(c)
suggest that \emph{it is important to measure beyond just the first
peak in the scattering pattern}. Second, it is essential that \emph{the
statistical errors in the data are low}; specifically, they should
be low enough that the modulation in the powder scattering pattern
is well resolved even at high values of $Q$. While these criteria
may be difficult to achieve with samples that are small or very highly
absorbing (particularly if the material in question also has a small
$S$), in most other cases they should be attainable using current-generation
diffuse-scattering diffractometers such as D7 (ILL) \cite{Stewart_2009_2}
or DNS (J\"ulich) \cite{Schweika_2001}. Refinement of experimental
data using SPINVERT is therefore likely to be both practical and informative---a
point we hope to demonstrate in the following section.

\section{Legacy data: Y$_{0.5}$Ca$_{0.5}$BaCo$_4$O$_7$ \label{sec:Legacy-data:-}}

As our final case study we demonstrate one of the key uses we envisage
for the SPINVERT program: analysis of legacy data. We consider spin
correlations in the compound Y$_{0.5}$Ca$_{0.5}$BaCo$_4$O$_7$
\cite{Schweika_2007}. Based on powder diffuse scattering data, this
compound was originally suggested to realise a kagome-lattice model
similar to that studied in Section~\ref{sec:Kagome-Heisenberg-magnet}
\cite{Schweika_2007}. However, subsequent single-crystal experiments
on the parent compound YBaCo$_4$O$_7$ suggested a different
picture \cite{Manuel_2009}. Here we attempt to reconcile these two
sets of results, by using the original powder data of Ref.~\cite{Schweika_2007}
to predict the single-crystal data obtained in Ref.~\cite{Manuel_2009}.

\begin{figure}
\begin{centering}
\includegraphics[scale=0.875]{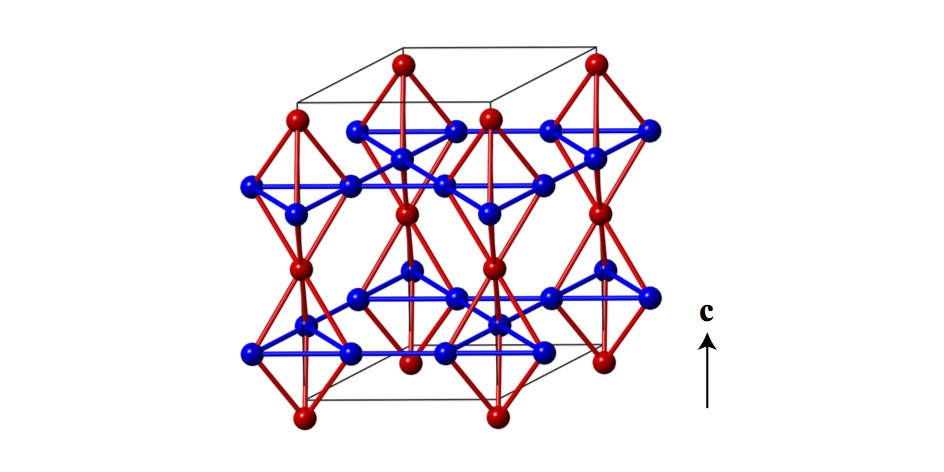}
\par\end{centering}

\caption{\label{fig:ybco_structure}Arrangement of magnetic Co ions in Y$_{0.5}$Ca$_0.5$Ba
Co$_4$O$_7$. The Co2 (Co$^{2+}$/Co$^{3+}$) sites
form kagome layers, shown in blue. The Co1 (Co$^{3+}$) sites form triangular
layers, shown in red. The structure can be described as a network
of corner-sharing trigonal bipyramids.}
\end{figure}

The crystal structure of Y$_{0.5}$Ca$_{0.5}$BaCo$_4$O$_7$
(hexagonal $P6_{3}mc$ \cite{Valldor_2006} or trigonal $P31c$ \cite{Stewart_2011})
has a unit cell containing eight magnetic Co ions, spread over two
inequivalent sites (Co1 and Co2). The Co1 sites form triangular layers
and the Co2 sites kagome layers, with the two kinds of layers alternating
along the $c$ axis, identifying the potential for spin frustration
[Fig.~\ref{fig:ybco_structure}]. The Co1 atoms are in the Co$^{3+}$
valence state, and the Co2 atoms are mixed-valence Co$^{3+}$
and Co$^{2+}$; both Co sites are tetrahedrally coordinated
by oxygen. The spin correlations in Y$_{0.5}$Ca$_{0.5}$BaCo$_4$O$_7$
were first studied in detail in Ref.~\cite{Schweika_2007}. In that
study, polarised-neutron powder diffraction data showed the absence
of periodic magnetic order to $T=1.2$\,K. It was considered that
this lack of magnetic order suggested that the kagome planes were
essentially decoupled. This was explained by postulating that the
Co1 (triangular lattice) ions are low-spin $S=0$, rather than high-spin
$S=2$ as might be expected. The experimental basis for this model
was that the magnetic moment obtained from fitting the neutron scattering
data was significantly lower than the calculated value assuming that
both Co sites were high-spin. However, several subsequent studies
have suggested a need for this conclusion to be revised. First, an
inelastic neutron scattering study \cite{Stewart_2011} shows that
a high-frequency component to the spin fluctuations is present, which
lies beyond the dynamic range of the measurements perfumed in Ref.~\cite{Schweika_2007}.
When this component to the scattering is taken into account, the expected
moment for a model with both Co sites high-spin is recovered. Second,
X-ray absorption spectra measurements for the parent compound, YBaCo$_4$O$_7$,
indicate that both Co sites are high-spin \cite{Hollmann_2009}. It
was also noted that in $T_d$ symmetry an $S=0$ state for Co requires
two spins to align parallel in a $t_2$ orbital, in disagreement
with Hund's first rule. To allow a $S=0$ state a very large distortion
from $T_d$ symmetry would be required, whereas the crystal structure,
as determined by neutron diffraction, shows only a small distortion
to $C_{3v}$ \cite{Stewart_2011}. 

Perhaps most significant is a study reporting single-crystal neutron
scattering measurements of the parent compound, YBaCo$_4$O$_7$
\cite{Manuel_2009}. Unlike Y$_{0.5}$Ca$_{0.5}$BaCo$_4$O$_7$,
oxygen-stoichiometric YBaCo$_4$O$_7$ undergoes a transition
to an antiferromagnetic state at $T_{\mathrm{N}}\simeq110$\,K \cite{Chapon_2006}.
Just above $T_{\mathrm{N}}$, at $T=130$\,K, broad diffuse features
were observed in the $(hk0)$ reciprocal-space
plane, indicating short-range magnetic order within the kagome planes.
Most interestingly, much sharper magnetic peaks were also observed
in the perpendicular scattering plane, identifying the presence of
longer-range correlations along the $c$ axis. These features were
successfully reproduced using a spin Hamiltonian in which both Co
sites are magnetic: it was found that Co1 spins align ferromagnetically
with a long correlation length, whereas the Co2 kagome spins remain
in a spin-liquid-like state. 

\begin{figure}
\begin{centering}
\includegraphics[scale=0.875]{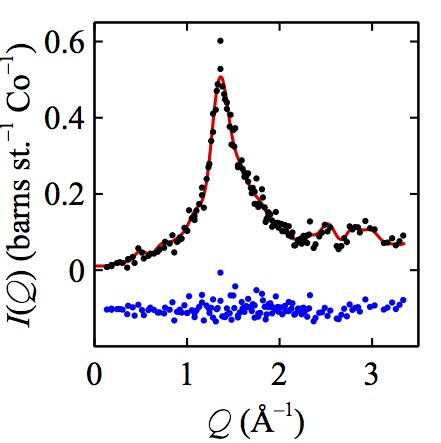}
\par\end{centering}

\caption{\label{fig:ybco_fit}Powder magnetic diffuse scattering data for Y$_{0.5}$Ca$_{0.5}$Ba
Co$_4$O$_7$ from Schweika et al. The
temperature of data collection for the powder data $T=1.2$ K. Experimental
data are shown as black circles, SPINVERT fit as a red line, and difference
(data--fit) as blue circles.}
\end{figure}

An outstanding question is then: do the spin correlations in Y$_{0.5}$Ca$_{0.5}$BaCo$_4$O$_7$
resemble the model of independent kagome planes of Co2 spins proposed
in Ref.~\cite{Schweika_2007}, or do they follow the model of ferromagnetic
correlations between Co1 spins identified in Ref.~\cite{Manuel_2009}
for YBaCo$_4$O$_7$? The lack of single-crystal samples
of Y$_{0.5}$Ca$_{0.5}$BaCo$_4$O$_7$ renders a direct answer
inaccessible. However, we will show that a reasonably conclusive answer
can nevertheless be obtained by employing SPINVERT refinement of the
original powder data of Ref.~\cite{Schweika_2007}. We note the existence
of a previous RMC study Y$_{0.5}$Ca$_{0.5}$BaCo$_4$O$_7$ using
different data and refinement software \cite{Stewart_2011}; our results
are in broad agreement with this study, but go beyond it by addressing
the three-dimensional nature of the spin correlations. We performed
SPINVERT refinements using a $10\times6\times6$ supercell of the
orthorhombic unit cell containing 5760 spins (crystallographic details
are taken from Ref.~\cite{Valldor_2006}). In our refinements, we
assume both Co sites are magnetic. As a starting point, we also take
the magnetic moments on both Co sites to be of equal magnitude; refinements
were also performed assuming relative spin lengths $S=2$ (for Co$^{3+}$)
and $S=3/2$ (for Co$^{2+}$), which gave similar results.
Refinements were performed for 300 proposed moves per spin, after
which time no significant improvement was observed in $\chi^{2}$,
and 16 independent refinements were performed in order to obtain good
statistics for scattering calculations. An intensity scale factor
was allowed to refine to fit the data, from which we estimate $S\simeq1.1$
(taking $g=2$), in good agreement with the results of \cite{Stewart_2011,Schweika_2007}.
The fit-to-data we obtain is shown in Fig.~\ref{fig:ybco_fit}. All
the features of the data are captured very well, apart from a slight
peak broadening in the fit due to the finite box size. 

\begin{figure}
\begin{centering}
\includegraphics[scale=0.875]{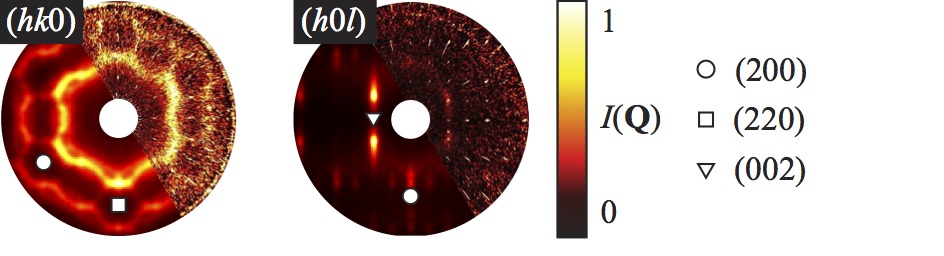}
\par\end{centering}

\caption{\label{fig:ybco_sc}Comparison of predicted magnetic diffuse scattering
for Y$_{0.5}$Ca$_{0.5}$BaCo$_4$O$_7$
with experimental data for YBaCo$_4$O$_7$. Two reciprocal-space
planes are shown, $(hk0)$ [left-hand image] and $(0kl)$
[right-hand image]. For each plane, the left panel shows the predicted
single-crystal scattering based on fitting the $T=1.2$\,K powder data
of Schweika \emph{et al.}\ for Y$_{0.5}$Ca$_{0.5}$BaCo$_4$O$_7$
\cite{Schweika_2007}, and the right panel shows the experimental
single-crystal data of Manuel \emph{et al.} for YBaCo$_4$O$_7$
at $T=130$\,K \cite{Manuel_2009}.}
\end{figure}

Having obtained spin configurations in agreement with the experimental
data, we proceed to characterise the nature of the spin correlations.
First we use the SPINDIFF program to calculate the single-crystal
scattering pattern from the SPINVERT spin configurations. This allows
a direct comparison between our prediction of the single-crystal scattering
for Y$_{0.5}$Ca$_{0.5}$BaCo$_4$O$_7$ and the experimental
single-crystal data for YBaCo$_4$O$_7$ from Ref.~\cite{Manuel_2009}.
This comparison is shown in Fig.~\ref{fig:ybco_sc}. Strikingly,
the experimental data and the SPINVERT prediction are almost identical.
Particularly significant is that the sharp reflections at $(\frac{1}{2}02)$
are present in both the SPINVERT refinement and the single-crystal
data. We note that the SPINVERT refinements are performed starting
from entirely random spin configurations, so these sharp features
cannot appear by chance: they must be required by the input powder
data. Therefore, a model of decoupled kagome planes with conventional
antiferromagnetic correlations (as considered in Section~\ref{sec:Kagome-Heisenberg-magnet})
is ruled out. Indeed, there are qualitative differences between the
powder diffuse scattering for such a model [Fig.~\ref{fig:kagome_fit}]
and the experimental powder data shown in Fig.~\ref{fig:ybco_fit}.

\begin{figure}
\begin{centering}
\includegraphics[scale=0.875]{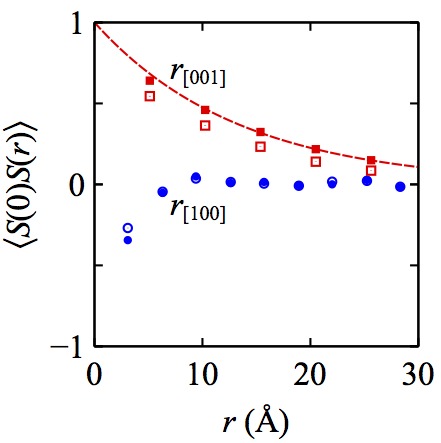}
\par\end{centering}

\caption{\label{fig:ybco_scf}Spin correlation function for Y$_{0.5}$Ca$_{0.5}$BaCo$_4$O$_7$
calculated from SPINVERT. Red squares show ferromagnetic correlations
along the $[001]$ real-space direction ($c$-axis). Blue
circles show antiferromagnetic correlations along $\langle 100\rangle $-type
directions within the kagome planes. Filled points are values obtained
from refinements assuming magnetic moments of equal length for both
Co sites; hollow points assume $S=2$ for Co$^{3+}$ and
$S=3/2$ for Co$^{2+}$.}
\end{figure}

Following Ref.~\cite{Manuel_2009} we plot the spin correlations
along two directions in real space. In Fig.~\ref{fig:ybco_scf},
we show correlations along hexagonal $[100]$-type directions
within the kagome plane (blue circles), and along the $[001]$
directions (red squares). The correlations along $[100]$
are antiferromagnetic and relatively weak. The correlations along
$[001]$ are always ferromagnetic, and are significantly
stronger, with exponential correlation length $\xi\approx13$~\AA.
In these essential features, the correlation functions we have shown
agree closely with the model for YBaCo$_4$O$_7$ (Fig.~3 of Ref.~\cite{Manuel_2009}). We note that a quantitative comparison
is not possible with the correlation functions shown in Ref.~\cite{Manuel_2009}:
their results were obtained from direct Monte Carlo simulations at
an arbitrary low temperature, and hence will show much stronger short-range
order than we observe for Y$_{0.5}$Ca$_{0.5}$BaCo$_4$O$_7$.
However, all the qualitative features, both in real and reciprocal
space, are in excellent agreement. 

In conclusion, we have found that the spin correlations in Y$_{0.5}$Ca$_{0.5}$BaCo$_4$O$_7$
closely resemble those in YBaCo$_4$O$_7$. The degree of
resemblance is remarkable, given that there is a difference in the
temperature of data collection of two orders of magnitude between
the single-crystal data of Ref.~\cite{Manuel_2009} and the powder
data of Ref.~\cite{Schweika_2007}. This result may be explained
by the substitution of Ca for Y in Y$_{0.5}$Ca$_{0.5}$BaCo$_4$O$_7$,
which introduces disorder into the exchange pathways, in effect reducing
the strength of the magnetic interactions. This disorder may also
help explain why Y$_{0.5}$Ca$_{0.5}$BaCo$_4$O$_7$ fails to
order magnetically, although the mechanism by which order is prevented
is not yet understood. Perhaps most importantly, the observation that
the key features of single-crystal data can be reproduced by fitting
\emph{powder} data for a real system presents several opportunities.
In the family of compounds based on YBaCo$_4$O$_7$, it
was recently suggested that chiral order may be present in the paramagnetic
phase \cite{Khalyavin_2012}: the appropriate correlation functions
are readily calculable using RMC techniques, so this intriguing possibility
could be in principle be investigated using SPINVERT. More generally,
we would argue that the initial uncertainty about the nature of the
spin correlations in Y$_{0.5}$Ca$_{0.5}$BaCo$_4$O$_7$ may
have arisen partially from a desire to impose a model onto experimental
data, rather than \emph{vice versa}. The essential advantage of the
reverse Monte Carlo approach is that the refinement is entirely data-driven,
so there is no requirement to think in terms of paradigms of frustrated
magnetism (\emph{e.g.}, independent kagome sheets) when performing
initial data analysis. Reality may be more complex than the models
we imagine; the RMC approach recognises this, and presents us with
results which are essentially model-independent.

\section*{Conclusions}

In this paper we have detailed the SPINVERT program for refinement
of paramagnetic powder scattering data using a reverse Monte Carlo
approach. This method is capable of reconstructing the three-dimensional
magnetic neutron scattering pattern from spherically-averaged data
\cite{Paddison_2012}. In addition, all the spin correlation functions
of the system are accessed. The RMC technique is primarily sensitive
to spin-pair correlations; however, in favourable cases we recover
information about higher-order spin correlations. We have not considered
here whether information on single-ion anisotropy can be obtained,
preferring to treat the anisotropy as a structural constraint. However,
we anticipate that the technique may also have a limited degree of
sensitivity to single-ion effects in systems with relatively low crystal
symmetry.

A key point about RMC is that it is entirely independent of a spin
Hamiltonian. This may be an advantage or a disadvantage: an advantage,
because it is not necessary to assume a form of the Hamiltonian to
model the spin correlations; a disadvantage, because it does not produce
such a microscopic model as output. Compared to other model-independent
techniques for analysis of diffuse scattering (such as simple curve
fitting), we believe that the RMC approach is superior in both the
quantity and accuracy of information it provides. Most importantly,
the crystal structure is a key ingredient in RMC refinements. Knowledge
of the crystal structure places strong constraints on the form of
the magnetic correlations, and it is this extra information allows
the three-dimensional scattering function $I\mathbf{(Q)}$
and spin correlation function $\langle\mathbf{S}\mathbf{(0)}\cdot\mathbf{S}(\mathbf{r})\rangle $
to be reconstructed. In fact, the key point here may not relate directly
to RMC at all: it is simply that the magnetic powder diffraction data---together
with knowledge of the the crystal structure---contain a great deal
of information. It seems very likely that other techniques could be
developed to access this information effectively; for example, it
is possible to fit interaction parameters directly to powder data
within a mean-field theory model (see, \emph{e.g.}, \cite{Yavorskii_2006,Skoulatos_2009}).

There are also certain disadvantages to the RMC approach. Most importantly,
RMC tends to produces the most disordered spin configuration compatible
with experimental data \cite{McGreevy_2001}. In practice, we have
not found the extent to which disorder is overestimated to be too
great, provided the spin anisotropy is specified for Ising or XY systems
\cite{Paddison_2012}. However, it is not possible to know in advance
the extent to which disorder will be created for a given dataset.
It is important always to remember that the success of RMC depends
entirely on the quality of the data with which it is provided. The
data should show the whole of the first broad peak as well as some
of the subsidiary peaks, and should ideally have excellent statistics.
The lowest possible background is desirable, and for this reason polarised-neutron
diffraction data have significant advantages \cite{Stewart_2009_2}.
We also anticipate that RMC may be less successful for crystal structures
which are complex---in the sense that they contain many radial separations
between pairs of atoms which nearly overlap---since the effect of
the limited real-space resolution imposed by $Q\mathrm{_{max}}$ will
then become more significant.

We hope that the SPINVERT program will be useful in several situations.
One possible use is the re-analysis of legacy data, as exemplified
in Section~\ref{sec:Legacy-data:-}. More importantly, when large
single crystal samples are currently unavailable, single crystal-like
information can be obtained using SPINVERT by fitting high-quality
powder data. A few topical examples should illustrate the potential
for discovery. For example, the $S=1/2$ kagome system kapellasite
was recently synthesised \cite{Colman_2008}, with measurements of
the bulk properties and inelastic neutron scattering suggesting a
quantum spin-liquid phase \cite{Faak_2012}. In the absence of single-crystal
samples, SPINVERT refinement of powder data would allow detailed characterisation
of these spin-liquid correlations. Second, ``the metallic spin ice''
system Pr$_2$Ir$_2$O$_7$ \cite{Nakatsuji_2006} exhibits
a range of seemingly unique properties, including a spontaneous Hall
effect in the paramagnetic phase suggestive of chiral order \cite{Machida_2010}.
Only small single crystals ($\sim$1\,mm$^3$) are reported
\cite{Machida_2007}, so SPINVERT refinement of powder diffraction
data would allow a valuable comparison of the predicted $I(\mathbf{Q})$
with microscopic models \cite{Ikeda_2008}. Furthermore, the ability
of SPINVERT to use a structural model containing non-magnetic impurities
would allow studies of canonical frustrated spin glasses such as SrCr$_{9x}$Ga$_{12-9x}$O$_{19}$
(SCGO) (see, \emph{e.g.}, \cite{Broholm_1990,Iida_2012}) and Ba$_2$Sn$_2$ZnGa$_{10-7x}$Cr$_{7x}$O$_{22}$
(BSZCGO) \cite{Hagemann_2001}: a model of the structural disorder
could be produced using, \emph{e.g.}, pair distribution function techniques
\cite{Young_2011}, and this structural model used as a starting-point
for SPINVERT refinement of magnetic diffraction data. In this way,
the impact of the structural disorder on the magnetic correlations
could be compared with theoretical models \cite{Sen_2011}. Finally,
in some cases it may even be advantageous to perform experiments on
powder samples: for example, for high-throughput studies, for highly-absorbing
samples \cite{Stewart_2004}, or when only small single crystals are
available. Consequently, we believe that the SPINVERT code---available
at \emph{http://spinvert.chem.ox.ac.uk}---will enable useful information
to be obtained for a much wider range of cooperative paramagnets than
is presently the case.

\ack{We are grateful to M. J. Cliffe,
M. J. Gutmann, P. Manuel, L. C. Chapon, D. A. Keen, and B. D. Rainford for valuable discussions.
JAMP and ALG gratefully acknowledge financial support from the STFC,
EPSRC (EP/G004528/2) and ERC (Ref: 279705).}

\section*{References}{}


\providecommand{\newblock}{}

\end{document}